\documentclass{article}
\usepackage{graphicx}
\usepackage{cite}

\begin{document}

\title{Exact Ultra Cold Neutrons' Energy Spectrum in Gravitational Quantum Mechanics}
\author{Pouria Pedram\thanks{p.pedram@srbiau.ac.ir}\\
{\small Department of Physics, Science and Research Branch, Islamic
Azad University, Tehran, Iran}}

\date{\today}
\maketitle \baselineskip 24pt

\begin{abstract}
We find exact energy eigenvalues and eigenfunctions of the quantum
bouncer in the  presence of the minimal length uncertainty and the
maximal momentum. This form of Generalized (Gravitational)
Uncertainty Principle (GUP) agrees with various theories of quantum
gravity and predicts a minimal length uncertainty proportional to
$\hbar\sqrt{\beta}$ and a maximal momentum proportional to
$1/\sqrt{\beta}$, where $\beta$ is the deformation parameter. We
also find the semiclassical energy spectrum and discuss the effects
of this GUP on the transition rate of the ultra cold neutrons in
gravitational spectrometers. Then, based on the Nesvizhevsky's
famous experiment, we obtain an upper bound on the dimensionless GUP
parameter.
\end{abstract}


\textit{Pacs}: {04.60.Bc}

\section{Introduction}
The existence of a minimal length uncertainty is one of the common
aspects of various theories of quantum gravity such as string
theory, loop quantum gravity, quantum geometry, and black-hole
physics. All these proposals predict a minimum measurable length of
the order of the Planck length
$\ell_{\mathrm{P}}=\sqrt{G\hbar/c^3}\approx 10^{-35}m$ where $G$ is
the Newton's gravitational constant. In recent years, many papers
have appeared in the literature to address the effects of the
minimal length on various physical systems in the framework of the
Generalized Uncertainty Principle (GUP) \cite{felder,12,main1}.

Recently, several approaches have been developed for testing the
effects of quantum gravity which range from astronomical
observations \cite{laser3,laser3-2} to table-top experiments
\cite{laser4}. We can mention the proposal by Amelino-Camelia and
Lammerzahl that explains puzzling observations of ultrahigh energy
cosmic rays in the framework of (quantum gravity) modified laws for
particle propagation \cite{laser1}. The usage of high intensity
Laser projects for quantum gravity phenomenology is also suggested
by Magueijo in the context of deformed special relativity
\cite{laser2}. Pikovski \emph{et al.} have proposed a direct
measurement scheme to experimentally test the existence of a minimal
length scale using a quantum optical ancillary system \cite{laser4}.
They probed possible deviations from ordinary quantum commutation
relation at the Planck scale within reach of current technology.
These progresses could shed light on possible detectable
Planck-scale effects.

On the other hand, Doubly Special Relativity (DSR) theories
essentially require the existence of a maximal momentum \cite{21}. A
GUP proposal which is consistent with these theories is discussed in
Refs.~\cite{main,pedram1}. Moreover, another GUP scenario which
agrees with the seminal proposal by Kempf, Mangano and Mann (KMM)
\cite{Kempf} to first order of the GUP parameter is recently
proposed in Refs.~\cite{pedramHigh}. The latter GUP model implies
the noncommutative geometry, i.e.~$[X_i,X_j]\ne 0$, and predicts
both a minimal length uncertainty and a maximal momentum
proportional to $\hbar\sqrt{\beta}$ and $1/\sqrt{\beta}$
respectively, where $\beta$ is the GUP parameter. Various problems
such as the free particle, particle in a box, harmonic oscillator,
black body radiation, cosmological constant, maximally localized
states, and hydrogen atom have been previously studied in this
framework \cite{pedramHigh,pedramH2}.

The quantum stationary solutions of the Schr\"odinger equation that
describes particles bouncing vertically and elastically above a
horizontal reflecting mirror in the Earth's gravitational field is
well-known theoretically. However, the experimental realization of
this phenomenon is so difficult due to two main reasons. First, for
the macroscopic objects the gravitational quantum effects are
negligible. Also, the electromagnetic interaction has the dominate
role for the charged particles. Thus, we need to perform the
experiment using neutral elementary particles with a long lifetime
such as neutrons. This famous experiment has been demonstrated a few
years ago by Nesvizhevsky et al. using a high precision neutron
gravitational spectrometer \cite{291}. Notice that, the perturbative
effects of the minimal length and/or maximal momentum on the quantum
bouncer spectrum are studied in Refs.~\cite{Brau,pedram1,NP}. In the
context of the KMM GUP where there exists just a minimal length
uncertainty this problem is exactly solved in Ref.~\cite{pedramInt}.

\section{The generalized uncertainty principle}
Consider the following deformed commutation relation that implies
both a minimal length uncertainty and a maximal observable momentum
\cite{pedramHigh}
\begin{eqnarray}\label{gupc}
[X,P]=\frac{i\hbar}{1-\beta P^2},
\end{eqnarray}
where $\beta=\beta_0/(M_{\mathrm{P}} c)^2$ is the GUP parameter,
$M_{\mathrm{P}}$ is the Planck mass and $\beta_0$ is the
dimensionless GUP parameter. Note that the maximal momentum is
introduced kinematically without breaking translation invariance. It
is worth mentioning that this invariance will be broken in the
presence of the minimum momentum uncertainty namely by introducing
position operator on the right hand side of the canonical
commutation relation (CCR) \cite{k1}. This is due to the fact that
the minimum momentum uncertainty can arise from curvature, as it is
shown in Ref. \cite{k2}. To avoid the singularity problem we need to
write out that rational term on the right hand side of
Eq.~(\ref{gupc}) as a geometric series. Then the singularity of the
denominator becomes a finite radius of convergence of the geometric
series. Otherwise, we face with the issue that the momentum can also
be larger than the cutoff. Basically, we get three irreducible
representations of the CCR, one with the momentum confined to the
desired range, one with the momentum confined to above and the other
with the momentum confined to below. A similar issue arises when a
momentum cutoff is introduced dynamically in the action \cite{k3}.

It is straightforward to check that the above commutation relation
results in the generalized uncertainty relation $ \Delta X \Delta P
\geq \frac{\hbar/2}{1-\beta \left[ (\Delta P)^2+\langle
P\rangle^2\right]}$ and
\begin{eqnarray}
(\Delta X)_{min}=\frac{3\sqrt{3}}{4}\hbar\sqrt{\beta},\hspace{1cm}
P_{\max}=\frac{1}{\sqrt{\beta}}.
\end{eqnarray}
In the momentum space representation, $X$ and $P$ can be written as
\begin{eqnarray}\label{rep1}
P \phi(p)&=& p\,\phi(p),\\
X\phi(p)&=& \frac{i\hbar}{1 - \beta
p^2}\partial_p\phi(p).\label{rep2}
\end{eqnarray}
Using the symmetricity condition for the position operator the
completeness relation and the scalar product are given by $
\langle\psi|\phi\rangle=\int_{-1/\sqrt{\beta}}^{+1/\sqrt{\beta}}\mathrm{d}p
\left(1-\beta p^2\right)\psi^{*}(p)\phi(p)$ and $ \langle
p|p'\rangle= \delta(p-p')/(1-\beta p^2)$. Also, the identity
operator is $\int_{-1/\sqrt{\beta}}^{+1/\sqrt{\beta}}\mathrm{d}p
\left(1-\beta p^2\right) |p\rangle \langle p|=1$ and the normalized
(unphysical) eigenfunctions of the position operator in momentum
space read $ \langle p|x\rangle =\frac{\sqrt{3\sqrt{\beta}}}{2}
\exp\left[\frac{-ix
p}{\hbar}\left(1-\frac{\beta}{3}p^2\right)\right], $ where $x$
denotes the eigenvalues of $X$ \cite{pedramHigh}.

\section{The quantum bouncer}
Now consider a particle of mass $m$ which is bouncing vertically and
elastically on a reflecting mirror. The potential is
\begin{eqnarray}
V(X)=\left\{
\begin{array}{cc}
mgX&\mathrm{for}\quad X>0,\\\\ \infty\, &\mathrm{for}\quad X\leq 0,
\end{array}
\right.
\end{eqnarray}
where $g$ is the acceleration in the Earth's gravitational field. In
the absence of GUP, i.e.~$\beta=0$, this problem is exactly solvable
and the solutions are given by the Airy functions. Also, the energy
spectrum correspond to the zeros of the Airy function. In the
semiclassical approximation, the Bohr-Sommerfeld formula gives the
approximate energy levels as $
E_n\simeq\left(\frac{9m}{8}\left[\pi\hbar
g\left(n-\frac{1}{4}\right)\right]^2\right)^{1/3} $ which provides
almost exact results even for the low quantum levels.

\begin{figure}
\begin{center}
\includegraphics[width=8cm]{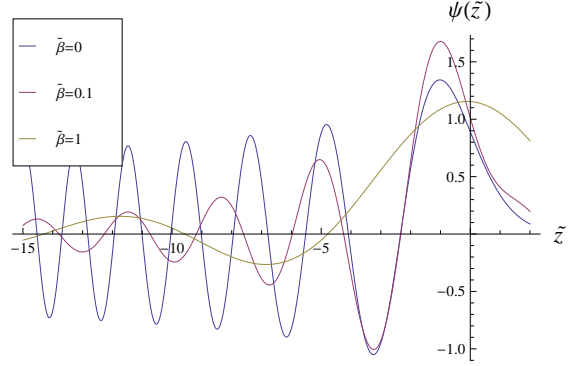}
\caption{\label{fig1}The wave function $\psi(\tilde{z})$ for
$\tilde\beta=\{0,0.1,1\}$.}
\end{center}
\end{figure}

The generalized Schr\"odinger equation now reads
\begin{eqnarray}
\frac{P^2}{2m}\phi+mgX \phi=E\,\phi.
\end{eqnarray}
In the momentum space and using Eqs.~(\ref{rep1},\ref{rep2}) we
obtain
\begin{eqnarray}
\frac{p^2}{2m}\phi(p)+\frac{i\hbar mg}{1-\beta p^2}
\phi'(p)=E\,\phi(p),
\end{eqnarray}
where prime denotes the derivative with respect to $p$. The solution
is
\begin{eqnarray}
\phi(p)=\phi_0 \exp\left[\frac{i\alpha
p}{\hbar}\left(\frac{p^2}{3}-2mE\right)\right]\exp\left[\frac{-i\alpha
\beta p^3}{\hbar}\left(\frac{p^2}{5}-\frac{2mE}{3}\right)\right],
\end{eqnarray}
where $\alpha^{-1}=2m^2g$. Now, using the identity operator, the
wave function in coordinate space is given by
\begin{eqnarray}
\psi(x)&=&\int_{-1/\sqrt{\beta}}^{+1/\sqrt{\beta}}
\left(1-\beta p^2\right) \langle x|p\rangle\phi(p)\mathrm{d}p,\nonumber\\
&=&\mathcal{A}\int_{-1/\sqrt{\beta}}^{+1/\sqrt{\beta}} \left(1-\beta
p^2\right) \exp\left[\frac{ix
p}{\hbar}\left(1-\frac{\beta}{3}p^2\right)\right]\phi(p)\mathrm{d}p.\hspace{.75cm}
\end{eqnarray}
So in terms of the new variable $z=x-\frac{E}{mg}$ we find
\begin{eqnarray}
\psi(z)=\mathcal{A}\int_{-1/\sqrt{\beta}}^{+1/\sqrt{\beta}}
\left(1-\beta p^2\right) \exp\left[\frac{i\alpha
p^3}{3\hbar}\left(1-\frac{3}{5}\beta p^2\right)\right]
\exp\left[\frac{i z
p}{\hbar}\left(1-\frac{\beta}{3}p^2\right)\right]\mathrm{d}p.
\end{eqnarray}
To proceed further, let us define the dimensionless variables
$\tilde{z}=z/l$, $\tilde{p}=p/k$, and $\tilde\beta=k^2\beta$, where
$l$ and $k$ have dimension of length and momentum, respectively, and
$kl=\hbar$. So the wave function reads
\begin{eqnarray}
\psi(\tilde{z})=\tilde{\mathcal{A}}\int_{-1/\sqrt{\tilde\beta}}^{+1/\sqrt{\tilde\beta}}
\left(1-\tilde\beta \tilde{p}^2\right) \exp\left[\frac{i\alpha
k^3}{3\hbar}\tilde{p}^3\left(1-\frac{3}{5}\tilde\beta
\tilde{p}^2\right)\right] \exp\left[i\tilde{z}
\tilde{p}\left(1-\frac{\tilde\beta}{3}\tilde{p}^2\right)\right]\mathrm{d}\tilde{p}.
\end{eqnarray}
Because of the Dirichlet boundary condition at $x=0$, the energy
eigenvalues are  given by $E_n=-mgl\tilde{z}_n$ where $\tilde{z}_n$
are given by $\psi(\tilde{z})\big|_{\tilde{z}_n}=0$. Now we set
$k=\left(\frac{\hbar}{\alpha}\right)^{1/3}=\left(2m^2g\hbar\right)^{1/3}$
and evaluate the above integral numerically. So we have
$l=\hbar^{2/3}\alpha^{1/3}=\left(\frac{\hbar^2}{2m^2g}\right)^{1/3}$
and the energy eigenvalues can be written as
$E_n=-\left(\frac{m\hbar^2g^2}{2}\right)^{1/3}\tilde{z}_n$, where in
the absence of GUP ($\tilde\beta=0$), $\tilde{z}_n$ are zeros of the
Airy function and we obtain the energy spectrum of the quantum
bouncer in ordinary quantum mechanics. Figure \ref{fig1} shows the
wave function $\psi(\tilde{z})$ for $\tilde{\beta}=\{0,0.1,1\}$
where its zeros correspond to the energy spectrum. Also, Table
\ref{tab1} shows the first ten quantized energies of the quantum
bouncer in the absence of GUP and in the presence of the minimal
length and maximal momentum. As it can be seen from the table, All
energy levels satisfy
\begin{eqnarray}
E_{\beta=0}<E_{\beta\ne0}.
\end{eqnarray}

\begin{table}
\centering
\begin{tabular}{ccc}\hline
$n$&$\tilde\beta=0$  & $\tilde\beta=1$ \\   \hline
1& 2.33811       & 4.77920   \\
2& 4.08795       & 9.57055   \\
3& 5.52056       & 14.3063   \\
4& 6.78671       & 19.0290   \\
5& 7.94413       & 23.7470   \\
6& 9.02265       & 28.4629   \\
7& 10.0402       & 33.1776   \\
8& 11.0085       & 37.8916   \\
9& 11.9360       & 42.6051   \\
10&12.8288       & 47.3197   \\\hline
\end{tabular}
\caption{\label{tab1}The first ten quantized energies of a bouncing
particle.}
\end{table}

To explain this fact, let us study the classical dynamics of a
particle $ \frac{dP}{dt}=\{P,H\}$ where the Poisson bracket in
classical mechanics corresponds to  the commutator in quantum
mechanics as $ \frac{1}{i\hbar}[\hat{A},\hat{B}]\Rightarrow\{A,B\}$.
For our case we find $ \frac{dP}{dt}=mg\{P,X\}=\frac{-mg}{1-\beta
P^2}=-mG(P)$ where $G(P)>g$ is the effective momentum-dependent
acceleration. So the increase in the acceleration implies the
increases in the energy spectrum in the GUP framework.

To distinguish the effect of this GUP from the one studied in
Ref.~\cite{pedramInt}, we should mention that because of the maximum
momentum, the number of the eigenstates are finite and the energy
spectrum consists of a maximum bound proportional to $\beta^{-1}$.
As we shall see, in semiclassical approximation, we derive
analytical relations for the number of states and the maximal energy
for the quantum bouncer in this GUP framework.

The observation of spontaneous decays of excited states in
Nesvizhevsky's experiment  can be considered as the manifestation of
the quantization of the quantum bouncer states in the presence of
the gravitational field \cite{30}. On the other hand, since the
energy eigenvalues of this quantum system increases in the presence
of the minimal length and the maximal momentum, the rate of these
decays will be changed as a Planck-scale effect. The transition
probability in the presence of the  generalized uncertainty
principle and in quadrupole approximation reads \cite{30}
\begin{eqnarray}
\Gamma_{k\longrightarrow
n}=\frac{4}{15}\frac{\omega^5_{kn}}{M_{\mathrm{P}}^2c^4}Q_{kn}^2,
\end{eqnarray}
where $\omega_{kn}=(E_k-E_n)/\hbar$ is the angular frequency of the
transition and
\begin{eqnarray}
Q_{kn}&=&m\langle
k|X^2|n\rangle\nonumber\\
&=&\frac{\displaystyle-m\hbar^2\int_{-1/\sqrt{\beta}}^{+1/\sqrt{\beta}}\mathrm{d}p\,
\phi^{*}(p)\frac{\mathrm{d}}{\mathrm{d}p}\frac{1}{1 - \beta
p^2}\frac{\mathrm{d}}{\mathrm{d}p}\phi(p)}{\displaystyle\int_{-1/\sqrt{\beta}}^{+1/\sqrt{\beta}}\mathrm{d}p
\left(1-\beta p^2\right)\phi^{*}(p)\phi(p)},
\end{eqnarray}
is the quantum quadrupole moment for the transition $k\rightarrow
n$. Although the modification of the  ordinary transition rate is
expected to be small, the effects of the generalized uncertainty
principle on the transition rate of ultra cold neutrons are hoped to
be found in the future experiments.

The energy spectrum can be also estimated using the semiclassical
scheme. In the classical domain, the Hamiltonian of the quantum
bouncer is
\begin{eqnarray}\label{sem1}
\frac{p^2}{2m}+\frac{mg x}{1-\beta p^2}=E,
\end{eqnarray}
and the Bohr-Sommerfeld quantization rule reads
\begin{eqnarray}
\oint p\,\mathrm{d}x=
\left(n-\frac{1}{4}\right)h,\hspace{1cm}n=1,2,\ldots.
\end{eqnarray}
Note that, since the wave function is exactly zero for  $x\leq0$,
but it can penetrate into the right hand classically forbidden
region, we need to add the term $-1/4$ to the right hand side of the
above equation. The validity of the WKB approximation for this
modified quantum mechanics in also discussed in
Ref.~\cite{pedramHigh}. Now, because of $\oint p\,\mathrm{d}x=-\oint
x\,\mathrm{d}p$ we have
\begin{eqnarray}
\oint
x\,\mathrm{d}p&=&\frac{1}{m^2g}\int_{\sqrt{2mE}}^0 \left(1-\beta p^2\right) \left(2mE-p^2\right)\,\mathrm{d}p,\nonumber\\
&=&-\frac{4\sqrt{2}}{3\sqrt{m}g}\left(E^{3/2}-\frac{2}{5}m\beta
E^{5/2}\right).
\end{eqnarray}
So we finally obtain
\begin{eqnarray}\label{root}
\epsilon^{3/2}_n-\frac{1}{5}\tilde\beta
\epsilon^{5/2}_n=\frac{3\pi}{2}\left(n-\frac{1}{4}\right),
\end{eqnarray}
where $\epsilon_n=E_n/\left(m\hbar^2g^2/2\right)^{1/3}$. The
zeroth-order solution coincides with the well-known semiclassical
result, i.e., $\epsilon^0_n=\left[\frac{3}{2}\pi
\left(n-1/4\right)\right]^{2/3}$. Moreover, to the first-order the
solution is $
\epsilon^1_n=\epsilon^0_n\left(1+\frac{2}{15}\tilde\beta\epsilon^0_n\right)
$, which as we have expected, it agrees with the result of
Ref.~\cite{pedramInt} to this order of approximation.

\begin{table}
\centering
\begin{tabular}{cccc}\hline
$n$&$\epsilon_n^{\mathrm{exact}}$  & $\epsilon_n^{\mathrm{SC}}$&
$|\frac{\Delta \epsilon_n}{\epsilon_n}|$
\\
\hline
1&  2.48894     & 2.48822 &  0.0003    \\
2&  4.89488     & 4.68787 &  0.0423    \\
3&  7.12806     & 6.82340 &  0.0427    \\
4&  9.29948     & 9.22021 &  0.0085    \\
5&  11.4435     & 12.8462 &  0.1226    \\\hline
\end{tabular}
\caption{\label{tab2}The exact and semiclassical energy levels of
the quantum bouncer for $\tilde\beta=0.2$.}
\end{table}

In Table \ref{tab2} we have reported the exact and semiclassical
energy spectrum of the quantum bouncer in the GUP framework for
$\tilde\beta=0.2$. As Eq.~(\ref{root}) shows, the existence of the
maximum energy is manifest in the semiclassical description of the
problem. The maximal energy is given by the condition
$\displaystyle\frac{\mathrm{d}}{\mathrm{d}\epsilon}\left[\epsilon^{3/2}-\frac{1}{5}\tilde\beta
\epsilon^{5/2}\right]=0$, i.e., $
\epsilon_{max}^{\mathrm{SC}}=\frac{3}{\tilde\beta}$ and $
n_{max}^{\mathrm{SC}}=\left\lfloor
\frac{4\sqrt{3}}{5\pi\tilde\beta^{3/2}}+\frac{1}{4}\right\rfloor $
where $\lfloor x\rfloor$ denotes the largest integer not greater
than $x$. However, the semiclassical number of states and maximal
energies differ considerably with the quantum mechanical results.
For instance, for $\tilde\beta=1$ we have
$\epsilon_{max}^{\mathrm{SC}}=3$ and $n_{max}^{\mathrm{SC}}=0$. But
its quantum mechanical values are
$\epsilon_{max}^{\mathrm{exact}}=61.4574$ and
$n_{max}^{\mathrm{exact}}=13$. Also, for $\tilde\beta=0.1$ we obtain
$\epsilon_{max}^{\mathrm{SC}}=30$, $n_{max}^{\mathrm{SC}}=14$, and
$\epsilon_{max}^{\mathrm{exact}}=15.3059$,
$n_{max}^{\mathrm{exact}}=9$.

Notice that, because of $\beta=\tilde\beta/k^2$, the dimensionless
GUP parameter $\beta_0=\left(M_{\mathrm{P}}c\right)^2\beta$ is given
by
\begin{eqnarray}
\beta_0=\frac{\left(M_{\mathrm{P}}c\right)^2}{\left(2m^2g\hbar\right)^{2/3}}\tilde\beta.
\end{eqnarray}
There are some upper bounds for $\beta_0$ in the literature based on
the accuracy of current experiments \cite{main1}. For instance, for
the Lamb shift the upper bound is given by $\beta_0<10^{36}$, and
for the Landau levels we have $\beta_0<10^{50}$. Also, for the
scanning tunneling microscope the current experiments imply
$\beta_0<10^{21}$. More accurate measurements can be used to test
these upper bounds or further tighten these values.

Now, based on the the exact quantum mechanical results and the
accuracy measured for the ground state in Nesvizhevsky's experiment
\cite{291}, the upper bound reads $\tilde\beta<0.2$ or
\begin{eqnarray}
\beta_0<3\times10^{58}.
\end{eqnarray}
The length scale corresponding to this dimensionless GUP parameter
is
$\ell_{\mathrm{GUP}}=\sqrt{\beta_0}\ell_{\mathrm{P}}\simeq10^{29}\ell_{\mathrm{P}}$.
The above upper bound is far weaker than that set by the electroweak
scale, namely $\beta_0\leq10^{34}$ \cite{main1}, but it is not
incompatible with it. Indeed, using more accurate results in the
future experiments, this upper bound is expected to get reduced by
several orders of magnitude. In this case, it represents a new and
intermediate length scale between the electroweak and the Planck
scale. Our results can be interpreted in two ways: (a) the
predictions of this GUP are too small to measure at present
($\beta_0\sim1$), or (b) there exists a new intermediate length
scale ($\beta_0\gg1$) that may appear in future experiments in the
Large Hadron Collider.

\section{Conclusions}
In this paper, we studied the problem of the quantum bouncer in the
presence of the minimal length uncertainty and the maximal momentum.
We solved the generalized Schr\"odinger equation in the momentum
space and found the exact energy eigenvalues and eigenfunctions. We
studied this problem semiclassically and showed that the number of
solutions are finite and the energy is bounded from above. These
effects are new and are absent in ordinary quantum mechanics and the
KMM GUP framework. Also, the transition probability of the ultra
cold neutrons in the presence of the generalized uncertainty
principle and in quadrupole approximation is obtained. We finally
found an upper bound on the dimensionless GUP parameter based on the
Nesvizhevsky's experiment.

\section*{Acknowledgement}
I am very grateful to Achim Kempf for insightful comments and
suggestions.

\end{document}